\documentclass[12pt]{iopart}
\usepackage{iopams}
\usepackage{amssymb}
\usepackage{mathrsfs}
\usepackage{graphicx}
\usepackage{bm}
\usepackage{epstopdf}
\usepackage{float}
\usepackage{braket}
\usepackage[colorlinks=true,citecolor=blue,linkcolor=blue,urlcolor=blue]{hyperref}

\begin{document}

\title[]{Electric- and magnetic-field dependence of the electronic and optical properties of phosphorene quantum dots}

\date{\today}

\author{L. L. Li$^{1,2}$, D. Moldovan$^1$, W. Xu$^{2,3}$ and F. M. Peeters$^{1}$}
\address{$^1$Department of Physics, University of Antwerp,
Groenenborgerlaan 171, B-2020 Antwerpen, Belgium}
\address{$^2$Key Laboratory of Materials Physics, Institute of
Solid State Physics, Chinese Academy of Sciences, Hefei 230031,
China}
\address{$^3$Department of Physics, Yunnan University, Kunming
650091, China}
\eads{\mailto{longlong.li@uantwerpen.be},
\mailto{dean.moldovan@uantwerpen.be},
\mailto{francois.peeters@uantwerpen.be}}

\begin{abstract}
Recently, black phosphorus quantum dots were fabricated experimentally. Motivated by these experiments, we theoretically investigate the electronic and optical properties of rectangular phosphorene quantum dots (RPQDs) in the presence of an in-plane electric field and a perpendicular magnetic field. The energy spectra and wave functions of RPQDs are obtained numerically using the tight-binding (TB) approach. We find edge states within the band gap of the RPQD which are well separated from the bulk states. In an undoped RPQD and for in-plane polarized light, due to the presence of well-defined edge states, we find three types of optical transitions which are between the bulk states, between the edge and bulk states, and between the edge states. The electric and magnetic fields influence the bulk-to-bulk, edge-to-bulk, and edge-to-edge transitions differently due to the different responses of bulk and edge states to these fields.
\end{abstract}

\maketitle

\section{Introduction}
Recently, two-dimensional (2D) black phosphorus (BP) has drawn a lot of attention from the research community. Bulk BP is a layered material in which the individual layers are stacked via weak van der Waals interactions. Single- and few-layer BP were experimentally fabricated from bulk BP \cite{LiL2014,LiuH2014}. Inside a single layer, each phosphorus atom is covalently bonded with three nearest phosphorus atoms to form a puckered honeycomb lattice. This unique lattice structure gives rise to anisotropic electronic and optical properties of 2D BP \cite{QiaoJ2014,XiaF2014}. Furthermore, compared to other known 2D materials such as graphene with zero band gap \cite{NovoselovK2005} and transition metal dichalcogenides (TMDs) with low carrier mobility \cite{WangQ2012}, 2D BP has the combined property of finite band gap ($\sim$1 eV) \cite{DasS2014} and high carrier mobility ($\sim$1000 cm$^2$/Vs) \cite{LiL2014}, which is crucial for practical applications in e.g. field-effect transistors.

At present, various interesting properties of 2D BP have been investigated theoretically and experimentally, such as strain-engineered band structure \cite{RodinA2014}, superior mechanical flexibility \cite{WeiQ2014}, tunable optical property \cite{LowT2014}, enhanced thermoelectric efficiency \cite{FeiR2014}, strong excitonic effect \cite{WangX2015}, nonlinear optical response \cite{LuS2015}, magneto-optical Hall effect \cite{TahirM2015}, and integer quantum Hall effect \cite{LiL2016}. BP nanoribbons (BPNRs) have also been studied, and their electronic, optical, thermal and transport properties were found to depend sensitively on the edge type and ribbon width \cite{CarvalhoA2014,PengX2014,TranV2014,ZhangJ2014,EzawaM2014, WuQ2015,SisakhtE2015,OstahieB2016}. A very striking property of BPNRs is the presence of topological edge states, which are well separated from the bulk states and can be greatly influenced by an external electric field \cite{EzawaM2014,SisakhtE2015}. However, compared to BP bulk and its nanoribbons, less attention has been paid to BP quantum dots (QDs). Most recently, BP QDs (BPQDs) have been successfully synthesized through chemical methods \cite{ZhangX2015,SunZ2015}. The obtained BPQDs have a lateral size of several nanometers and a thickness of few layers. Theoretical studies have also been carried out to investigate the electronic and optical properties of monolayer BPQDs or phosphorene QDs (PQDs) \cite{ZhangR2015,NiuX2016}. Some interesting results were obtained, such as unconventional edge states in PQDs \cite{ZhangR2015} and anomalous size-dependent optical properties of PQDs \cite{NiuX2016}. In particular, it has been shown \cite{ZhangR2015} that edge states appear stably in the band gap of the PQD regardless of its geometric shape and edge structure due to the anisotropic electron hopping in the system. As is known, from a device-application point of view, it is more convenient and efficient to tune the electronic and optical properties of a QD material by external electric and/or magnetic fields than by its geometric parameters such as size and shape. Based on this viewpoint, in the present work we investigate the electronic and optical properties of PQDs under external electric and magnetic fields. We employ the tight-binding (TB) method to study how these external fields influence the electronic and optical properties of PQDs.

Like modeling graphene QDs \cite{ZareniaM2011}, one has to take into account the effects of geometric shape and edge type when modeling PQDs using the TB method. The geometric shapes of experimentally fabricated BPQDs \cite{ZhangX2015} are rectangular-like \cite{NiuX2016} but there is no experimental identification of the edge types in such nanostructures. Because of this unknown, the authors in previous theoretical works \cite{ZhangR2015,NiuX2016} considered regular armchair and zigzag edges in PQDs with rectangular shapes. Although the edge types will probably be more complex in realistic BPQDs, this consideration is reasonable for theoretical modeling since the armchair and zigzag edges have been demonstrated to be chemically stable in BPNRs \cite{DasP2016}. Based on the above statements, in the present work we consider rectangular PQDs (RPQDs) with armchair and zigzag terminations.

This paper is organized as follows. In section II, we present the theoretical model for calculating the electronic and optical properties of RPQDs in the presence of electric and magnetic fields. In section III, we present and discuss the effects of electric and magnetic fields on the electronic and optical properties of RPQDs. Finally, we conclude our results with a summary in section IV.

\begin{figure}[htbp]
\centering
\includegraphics[width=8.9cm]{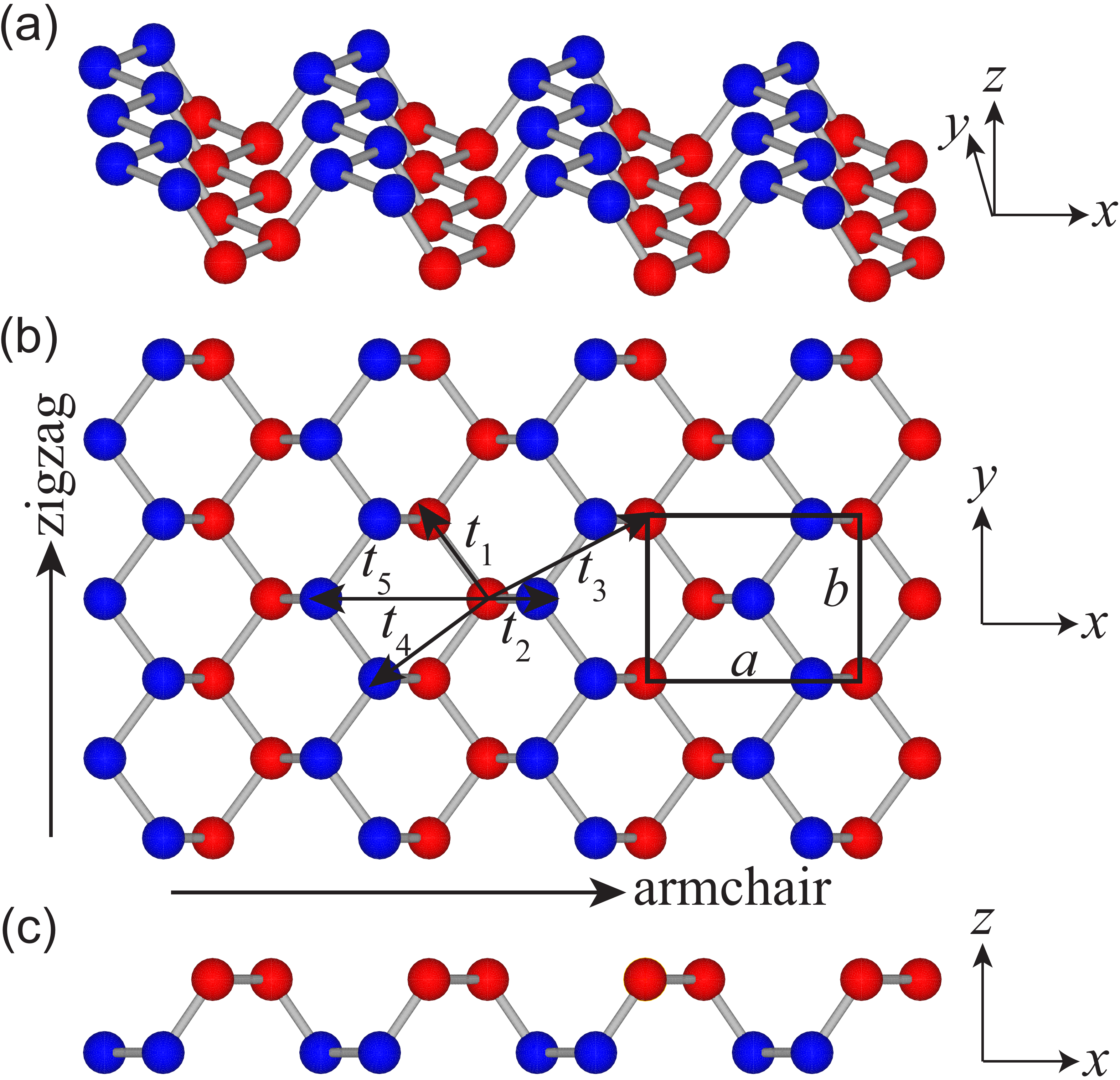}
\caption{Lattice structure of phosphorene: (a) 3D view, (b) top view and (c) side view. The unit cell consists of four inequivalent phosphorus atoms with two of them labeled by red solid circles and the other two labeled by blue solid circles. The symbols $t_i$ ($i=1, 2, ..., 5$) denote five hopping energies used in the tight-binding Hamiltonian of phosphorene. $a$ and $b$ are the lengths of the unit cell along the $x$ (armchair) and $y$ (zigzag) directions, respectively.} \label{Fig1}
\end{figure}

\section{Model and Theory}

The lattice structure of phosphorene is shown in Fig. 1. As can be seen, phosphorene has a puckered honeycomb lattice with unit-cell lengths $(a, b)=(0.443, 0.327)$ {nm} \cite{EzawaM2014}, and due to the puckered lattice structure there are four inequivalent phosphorus atoms in a unit cell. The TB Hamiltonian proposed for phosphorene is given by \cite{RudenkoA2014}
\begin{equation}\label{e1}
H=\sum_{i}\varepsilon_ic_i^{\dag}c_i+\sum_{i,j}t_{ij}c_i^{\dag}c_j,
\end{equation}
where the summation runs over the lattice sites of phosphorene,
$\varepsilon_i$ is the on-site energy of the electron at site $i$, $t_{ij}$ is the hopping energy between the $i^{th}$ and $j^{th}$ sites, and $c_i^{\dag}$ ($c_j$) is the creation (annihilation) operator of the electron at site $i$ ($j$). It has been shown \cite{RudenkoA2014} that it is sufficient to take five hopping energies to describe the band structure of phosphorene, as illustrated in Fig. 1. These five hopping energies are given by $t_1=-1.220$ eV, $t_2=3.665$ eV, $t_3=-0.205$ eV, $t_4=-0.105$ eV,
and $t_5=-0.055$ eV \cite{RudenkoA2014}. The on-site energies are taken as $\varepsilon=0$ for all the lattice sites. In the present work, we employ this TB model to study the electronic and optical properties of RPQDs with armchair and zigzag edges under external electric and magnetic fields.

We consider a RPQD placed in the $(x,y)$ plane (see Fig. 1), an external electric field applied along one of the in-plane directions (e.g. the $x$ or $y$ direction), and an external magnetic field applied perpendicular to the $(x,y)$ plane (i.e. the $z$ direction). When an in-plane electric field is applied to the RPQD, the on-site energy in the original TB Hamiltonian (1) should be modified by adding the electric potential term $-e\textbf{F}\cdot\textbf{r}_i$ with $e$ being the elementary charge, $\textbf{F}=(F_x,F_y)$ the in-plane electric field vector, and $\textbf{r}_i=(x_i,y_i)$ the in-plane position vector at site $i$. On the other hand, when a perpendicular magnetic field is applied to the system, the hopping energy in the original TB Hamiltonian (1) should be modified by multiplying with a phase factor determined by the magnetic flux, which is achieved via the so-called Peierls substitution:
\begin{equation}\label{e2}
t_{ij}\rightarrow t_{ij}\exp\Big(i\frac{2\pi e}{h}\int_{\textbf{r}_i}^{\textbf{r}_j}\textbf{A}\cdot d\textbf{l}\Big),
\end{equation}
with $h$ being the Planck constant and $\textbf{A}$ the vector potential induced by the magnetic field. In the Landau gauge, the magnetic vector potential can be written as $\textbf{A}=(0,Bx,0)$ with $B$ being the magnetic field strength. The magnetic flux threading a plaquette is defined as $\Phi=Bab$ in units of the flux quantum $\Phi_0=h/e$.

The energy levels and wave functions in the RPQD are obtained by diagonalizing the TB Hamiltonian matrix numerically. All numerical TB calculations are performed using the recently developed \textit{Pybinding} package \cite{MoldovanD2016}, and the obtained results are then imported into the calculations of the electronic density of states (DOS) and the optical absorption spectrum. The electronic DOS of a QD system is the sum of a series of delta functions, which can be numerically calculated with a Gaussian broadening as
\begin{equation}\label{e3}
D(E)=\frac{1}{\sqrt{2\pi\Gamma^2}}\sum_{n}\exp\Big[\frac{-(E-E_n )^2}{2\Gamma^2}\Big],
\end{equation}
where $\Gamma$ is the broadening factor and $E_n$ is the energy level for the $n$th eigenstate. The dipole matrix element for the transition from the initial state $\ket{i}$ to the final state $\ket{j}$ is given by $\textbf{M}_{ij}=\bra{j}\textbf{r}\ket{i}$, where light is assume to be polarized in the $(x,y)$ plane of the RPQD. The optical absorption is then calculated as $A(\hbar\omega)=\sum_{i,j}A_{ij}(\hbar\omega)$, where the summation is over all the possible dipole transitions and $A_{ij}(\hbar\omega)$ is given by \cite{LeeS2002,AbdelsalamH2016}
\begin{equation}\label{e4}
A_{ij}(\hbar\omega)=(E_j-E_i)\big|\boldsymbol{\epsilon}\cdot \textbf{M}_{ij}\big|^2\delta(E_i-E_j+\hbar\omega),
\end{equation}
where $E_i$ ($E_j$) is the energy level for the initial (final) state $\ket{i}$ ($\ket{j}$), $\boldsymbol{\epsilon}$ is the light polarization vector, and $\omega$ is the angular frequency of light. In the numerical calculation of $A(\hbar\omega)$, we use the same Gaussian broadening [Eq. (3)] for the delta function in Eq. (4).

\section{Single-particle spectrum: electric- and magnetic-field dependence}

In this part, we present and discuss our numerical results for the electronic and optical properties of RPQDs with armchair and zigzag edges in the presence of electric and magnetic fields. For convenience, we denote $L$ and $W$ as the side lengths of the RPQD along the armchair and zigzag directions, i.e., the $x$ and $y$ directions, respectively. The side lengths considered in the present work are taken as several nanometers which are comparable to those realized experimentally \cite{ZhangX2015,SunZ2015}.

\begin{figure}[htb]
\centering
\includegraphics[width=11.5cm]{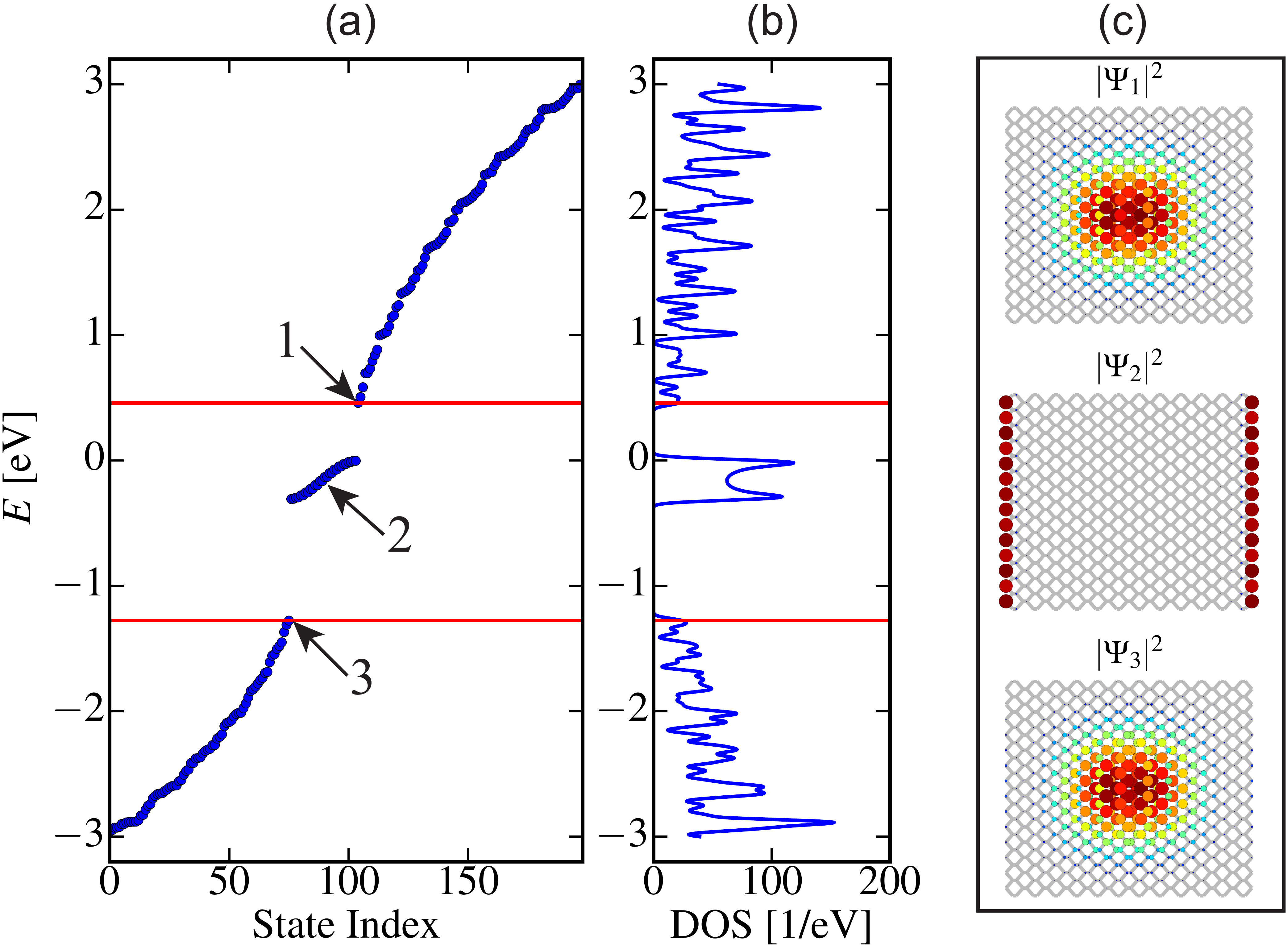}
\caption{(a) Energy levels, (b) density of states (DOS), and (c) wave functions of a RPQD with side lengths $(L, W)=(5.3, 4.1)$ nm at zero electric and magnetic fields. The wave functions in (c) are shown for the electronic states labeled by the numbers 1, 2 and 3 in (a). The region between the two red solid lines represents the band gap of the RPQD. In the DOS calculation, a broadening factor is taken of $\Gamma=0.02$ eV.} \label{Fig2}
\end{figure}

In Fig. 2, we show the energy levels, electronic DOS, and wave functions in a RPQD with side lengths $(L, W)=(5.3, 4.1)$ nm. Here, the region between the two red solid lines denotes the band gap of the RPQD. As can be seen, the edge states appear in the band gap of the RPQD and they are well separated from the bulk states [see Fig. 2(a)]. This means that these edge states are well-defined in the RPQD. The electronic DOS exhibits a two-peak structure within the band gap corresponding to the edge states while a many-peak structure outside the band gap to the bulk states [see Fig. 2(b)]. The two-peak structure for the edge states is caused by different sublattice contributions at the zigzag boundaries. The many-peak structure for the bulk states is caused by the discrete energy levels due to the usual quantum confinement effect. Here, we call the edge and bulk states according to their wave-function properties. To show such properties, we take three typical points labeled by the numbers 1, 2, and 3 in Fig. 2(a) and plot the squared wave functions for such three points in Fig. 2(c). It is clear that the edge states are only localized at the zigzag boundaries while the bulk states are mainly distributed in the central part of the RPQD. The zigzag edge states obtained here are not peculiar to phosphorene nanostructures, which can also occur in graphene and MoS$_2$ nanostructures as predicted by first-principle calculations \cite{Son2006,Bollinger2001}. We also note from our numerical calculations that the number of edge states is equal to the number of phosphorus atoms at the zigzag boundaries. The results shown in Fig. 2 indicate that edge states appear as mid-gap states in a RPQD with armchair and zigzag boundaries.

\begin{figure}[htb]
\centering
\includegraphics[width=12.5cm]{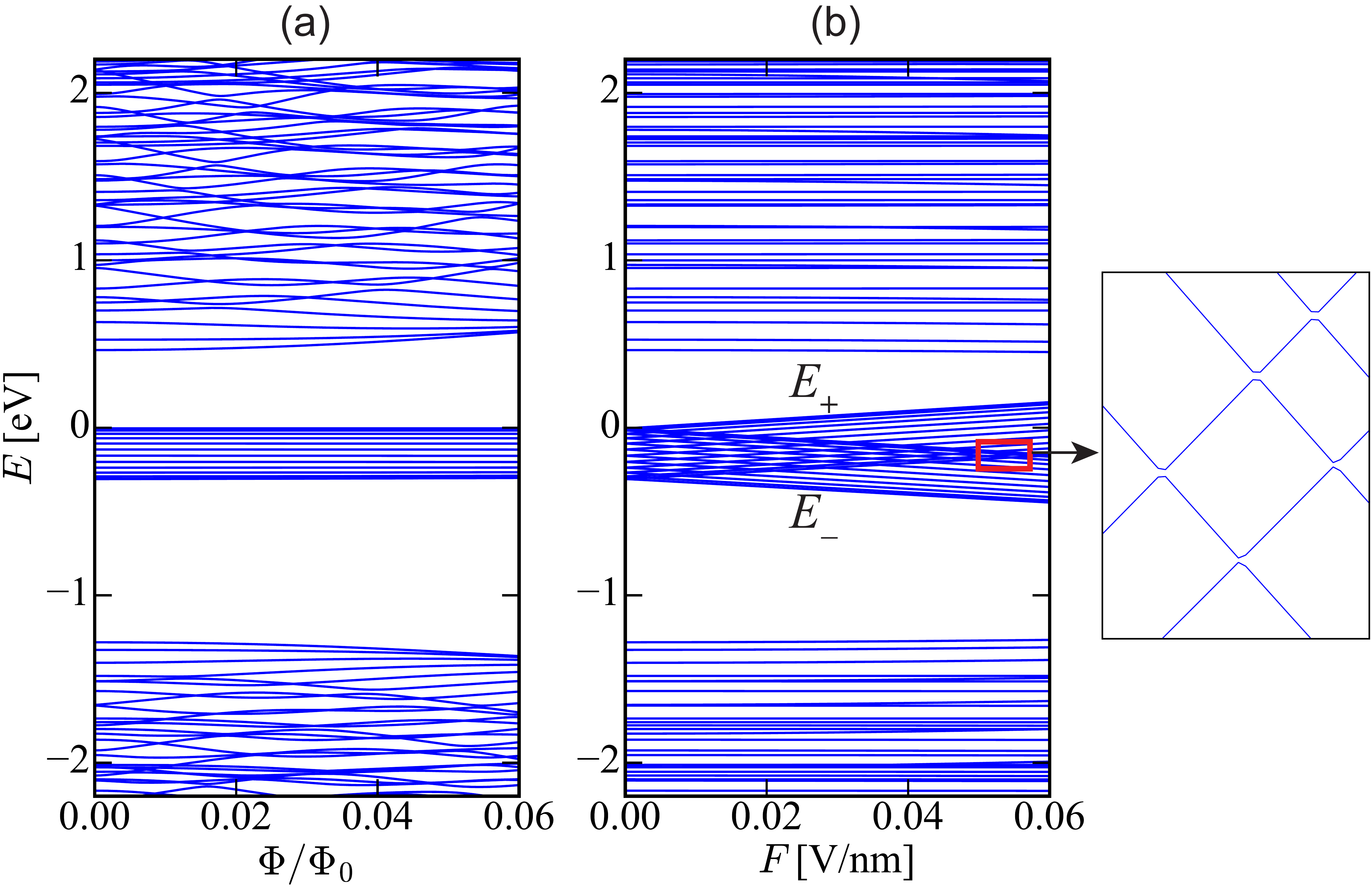}
\caption{Energy spectra of the same RPQD as in Fig. 2 in the presence of (a) perpendicular magnetic field and (b) in-plane electric field (applied along the $x$ direction). Here, $E_{+}$ and $E_{-}$ label the two split edge bands in the presence of an in-plane electric field. The red rectangular region in (b) is enlarged to show the anticrossings between the $E_{+}$ and $E_{-}$ bands. } \label{Fig1}
\end{figure}

In Fig. 3, we show the energy spectra of the same RPQD as in Fig. 2 in the presence of (a) perpendicular magnetic field and (b) in-plane electric field (applied along the $x$ direction). As can be seen, in the presence of magnetic field and with increasing magnetic flux, a nearly flat band is formed by the mid-gap edge states in the RPQD. The energy levels in this edge band are two-fold degenerate due to the presence of two identical zigzag boundaries. These edge levels are almost unaffected by the magnetic field, which is a consequence of the strong localized nature of edge states. However, the energy levels of the bulk states in the RPQD are non-degenerate due to the intrinsic asymmetry caused by the anisotropic boundaries, i.e., the armchair and zigzag boundaries along the $x$ and $y$ directions, respectively. These bulk levels correspond to the so-called Fock-Darwin states due to the competing geometric and magnetic confinements, and they will eventually approach the Landau levels as the magnetic flux further increases due to the dominant magnetic confinement.

On the other hand, in the presence of an electric field, the degeneracy of the edge levels is lifted due to the broken spatial inversion symmetry induced by the electric field. As a result, the single edge band at zero electric field is now split into two bands. One of these two bands (labeled as $E_{+}$) increases with increasing electric field and the other band (labeled as $E_{-}$) decreases with increasing electric field, which leads to anticrossings of the energy levels in the $E_{+}$ and $E_{-}$ bands. This feature can be understood within perturbation theory as follows. We denote the edge-state Hamiltonian in the presence of an electric field applied along the $x$ direction as $H_{F}=H_0-eFx$ with $H_0$ being the zero-field Hamiltonian and $F$ being the electric field strength. And the electric-field-induced part is treated as a perturbation. Since the edge levels are two-fold degenerate at zero electric field, we need to apply degenerate perturbation theory. Denoting $E_1$ and $E_2$ ($\ket{1}$ and $\ket{2}$) as two degenerate eigenenergies (corresponding eigenstates) of the unperturbed Hamiltonian $H_0$, we have $H_0\ket{1}=E_1\ket{1}$ and $H_0\ket{2}=E_2\ket{2}$. Within degenerate perturbation theory, the matrix elements of the perturbed Hamiltonian $H$ can be obtained in a degenerate basis set composed of unperturbed eigenstates $\{\ket{1}, \ket{2}\}$ as
\begin{equation}\label{e5}
[H_{F}]=\left[\begin{array}{cc}
E_0 \ \ &  \Delta \\
\Delta^{\dag} \ \ &  E_0 \\
\end{array}\right],
\end{equation}
where $E_0=E_1=E_2$ due to the degeneracy and $\Delta=-\bra{1}eFx\ket{2} \sim -eFL$. Diagonalizing this Hamiltonian matrix, we find that the energy levels exhibit a Stark shift under the electric-field perturbation as $E_{\pm}=E_0\pm|\Delta|$. Thus, the $E_{+}$ ($E_{-}$) band increases (decreases) linearly with increasing electric-field strength, as shown in Fig. 3(b). The up- and down-going behaviors of the $E_{+}$ and $E_{-}$ bands initially lead to the crossings of the energy levels but the interactions between the degenerate levels at the crossing points finally give rise to anticrossings, as shown by the zoomed plot in Fig. 3(b). However, compared to the edge states, the bulk states shown here are almost unaffected by the electric field due to the strong QD confinement.

\section{Optical absorption}

The optical absorption spectra of RPQDs in the presence of electric and magnetic fields are calculated with a constant broadening factor $\Gamma=0.01$ eV. This is in fact a relatively crude approximation because the realistic value of $\Gamma$ should be in principle calculated by considering both field-dependent electronic states and various scattering mechanisms. However, the main features of the absorption spectrum with this approximation are maintained and it has been still widely used in the literature \cite{HuangY2008,KoshinoM2008,TabertC2013}. Furthermore, in the present work we only consider the optical properties of undoped RPQDs, where the chemical potential is assumed to be $E_F=0$ eV and the temperature is taken as $T=0$ K, which implies that only optical transitions between the occupied states below $E_F$ and unoccupied states above $E_F$ are possible.

\begin{figure}[htbp]
\centering
\includegraphics[width=8.9cm]{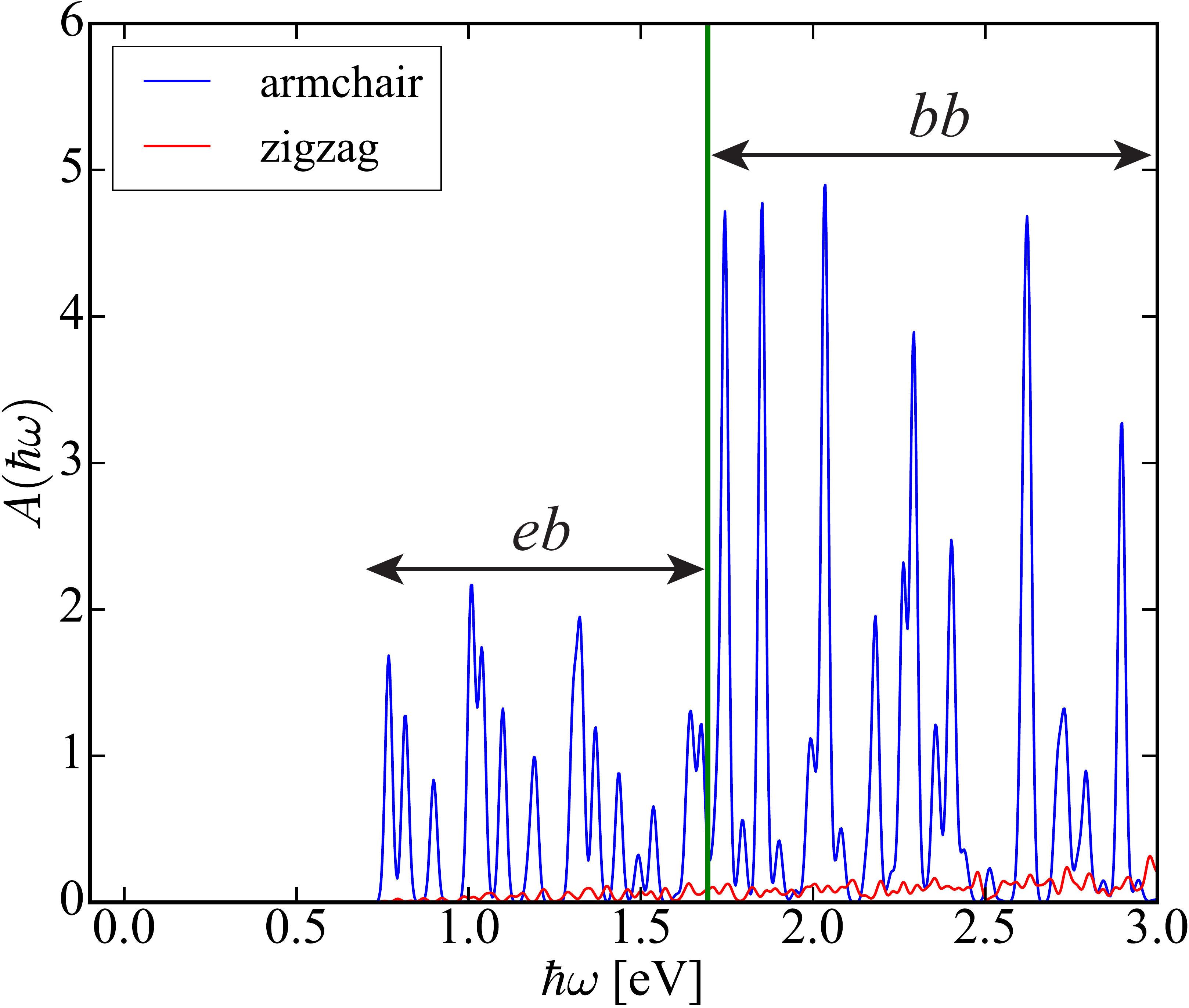}
\caption{Optical absorption spectra (arbitrary units) of the same RPQD as in Fig. 2 at zero electric and magnetic fields for different light polarizations as indicated. Here, the green solid line represents the band gap energy of the RPQD, and the characters $bb$ and $eb$ denote the bulk-to-bulk and edge-to-bulk transitions, respectively.} \label{Fig1}
\end{figure}

Now we give a brief analysis on the possible optical transitions from the view point of the energy spectrum. As shown in Fig. 3, due to the presence of edge states within the band gap of the RPQD, it is expected that optical transitions are possible between the bulk and bulk states, between the edge and bulk states, and between the edge and edge states.

In Fig. 4, we show the optical absorption spectrum of the same RPQD as in Fig. 2 in the absence of electric and magnetic fields for different light polarizations as indicated. Here, the green solid line represents the band gap energy of the RPQD. As can be seen, above the band gap, the optical absorption is mainly induced by those transitions from the bulk (hole) states to bulk (electron) states (the bulk-to-bulk transitions labeled $bb$ in the figure), while below the band gap, it is mainly induced those transitions from the edge states to bulk (electron) states (the edge-to-bulk transitions labeled $eb$ in the figure). The absorption induced by light polarized along the armchair direction (i.e. the $x$ direction) is much stronger than that induced by light polarized along the zigzag direction (i.e. the $y$ direction). The latter is almost insignificant in the given range of photon energies. This indicates that the RPQD can absorb (transmit) light polarized along the armchair (zigzag) direction. The polarization selective absorption is caused by highly anisotropic dipole transition strengths for light polarizations along the armchair and zigzag directions, which can be numerically determined by calculating the squared dipole matrix elements $|\bra{f}x\ket{i}|^2$ and $|\bra{f}y\ket{i}|^2$ with $\ket{i}$ and $\ket{f}$ being the initial and final states, respectively. Our numerical results indicate that $|\bra{f}x\ket{i}|^2$ is much larger than $|\bra{f}y\ket{i}|^2$. This is why we observe polarization sensitive absorption and a large linear dichroism in the RPQD. The linear dichroism of the optical absorption was also predicted for bulk phosphorene \cite{YuanS2015}.

As mentioned above, there are possibly three types of optical transitions in the RPQD. In the following, we will examine these possible optical transitions under electric and/or magnetic fields. To proceed, we define an averaged optical absorption $A_{ave}(\hbar\omega)=[A_{ac}(\hbar\omega)+A_{zz}(\hbar\omega)]/2$ with $A_{ac}(\hbar\omega)$ and $A_{zz}(\hbar\omega)$ being the optical absorptions induced by light polarized along the armchair and zigzag directions, respectively.

\begin{figure}[htbp]
\centering
\includegraphics[width=12.5cm]{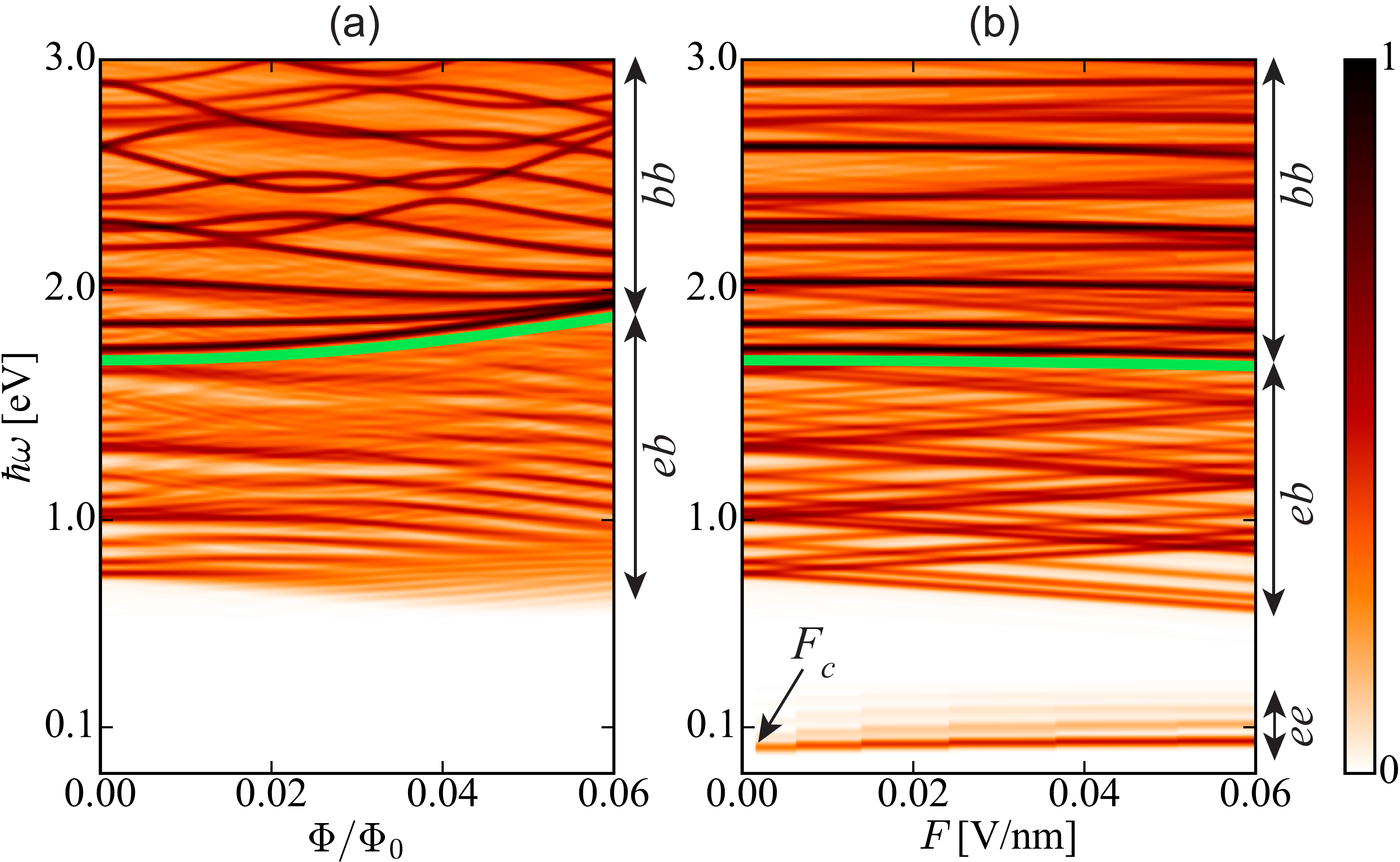}
\caption{Contour plots of the normalized optical absorption spectra of the same RPQD as in Fig. 2 in the presence of (a) perpendicular magnetic field and (b) in-plane electric field (applied along the $x$ direction). Here, the green solid line represents the optical gap of the RPQD, the characters $bb$, $eb$, and $ee$ denote the bulk-to-bulk, edge-to-bulk, and edge-to-edge transitions, respectively, $F_c\simeq0.001$ V/nm is a critical electric field, and for an electric field of $F>F_c$, the edge-to-edge transitions are activated. The white and black colors stand for the lowest and highest absorption intensities, respectively.} \label{Fig1}
\end{figure}

In Fig. 5, we show the contour plots of $A_{ave}(\hbar\omega)$ of the same RPQD as in Fig. 2 in the presence of (a) perpendicular magnetic field and (b) in-plane electric field (applied along the $x$ direction). Here, the optical absorption intensity is normalized with respect to its maximum value. The green solid line in each panel represents the optical gap which is defined as the photon energy that is equal to the global energy gap of the system. This global gap changes with electric and magnetic fields significantly (see Fig. 3), and it becomes the band gap of the RPQD at zero electric and magnetic fields, as shown in Fig. 4. As can be seen, the absorption lines vary with the magnetic and electric fields in different ways. For example, they vary linearly with electric field but nonlinearly with magnetic field. Such a difference is mainly caused by the different responses of bulk and edge states to the electric and magnetic fields, which are manifested in the corresponding energy spectra shown in Fig. 3. We observe two (three) types of optical transitions under magnetic (electric) fields: the first comes from the bulk-to-bulk transitions above the optical gap (labeled by $bb$ in the figure); the second from the edge-to-bulk transitions below the optical gap (labeled by $eb$ in the figure); and the third from the edge states themselves (the edge-to-edge transitions labeled by $ee$ in the figure) which is far below the optical gap. This peculiar optical absorption is due to the presence of edge states within the band gap of the RPQD (see the energy spectra shown in Fig. 3). The strong absorption lines above the optical gap are mainly induced by the bulk-to-bulk transitions for light polarization along the armchair direction, while the weak absorption lines below the optical gap are mainly induced by the edge-to-bulk transitions for the same light polarization. This is because the bulk and edge states are spatially separated: the former are mainly located around the center of the RPQD while the latter are mainly localized at the boundaries of the RPQD. When the electric field strength exceeds a critical value $F_c\simeq0.001$ V/nm, addition absorption lines emerge far below the optical gap (at photon energies around 0.1 eV), which are mainly induced by the edge-to-edge transitions for light polarization along the zigzag direction. This is because the electric field splits the single edge band into the two bands, in which some energy levels are above the Fermi energy while others are below the Fermi energy [see Fig. 3(b)]. Moreover, the absorption intensity induced by the edge-to-edge transitions increases with increasing electric field because more transition channels can be opened as the electric field increases [see Fig. 3(b)]. The distinct optical-absorption features induced by the bulk-to-bulk and edge-to-bulk transitions can be used to determine the band gap of the RPQD.

\begin{figure}[htbp]
\begin{center}
\includegraphics[width=8.5cm]{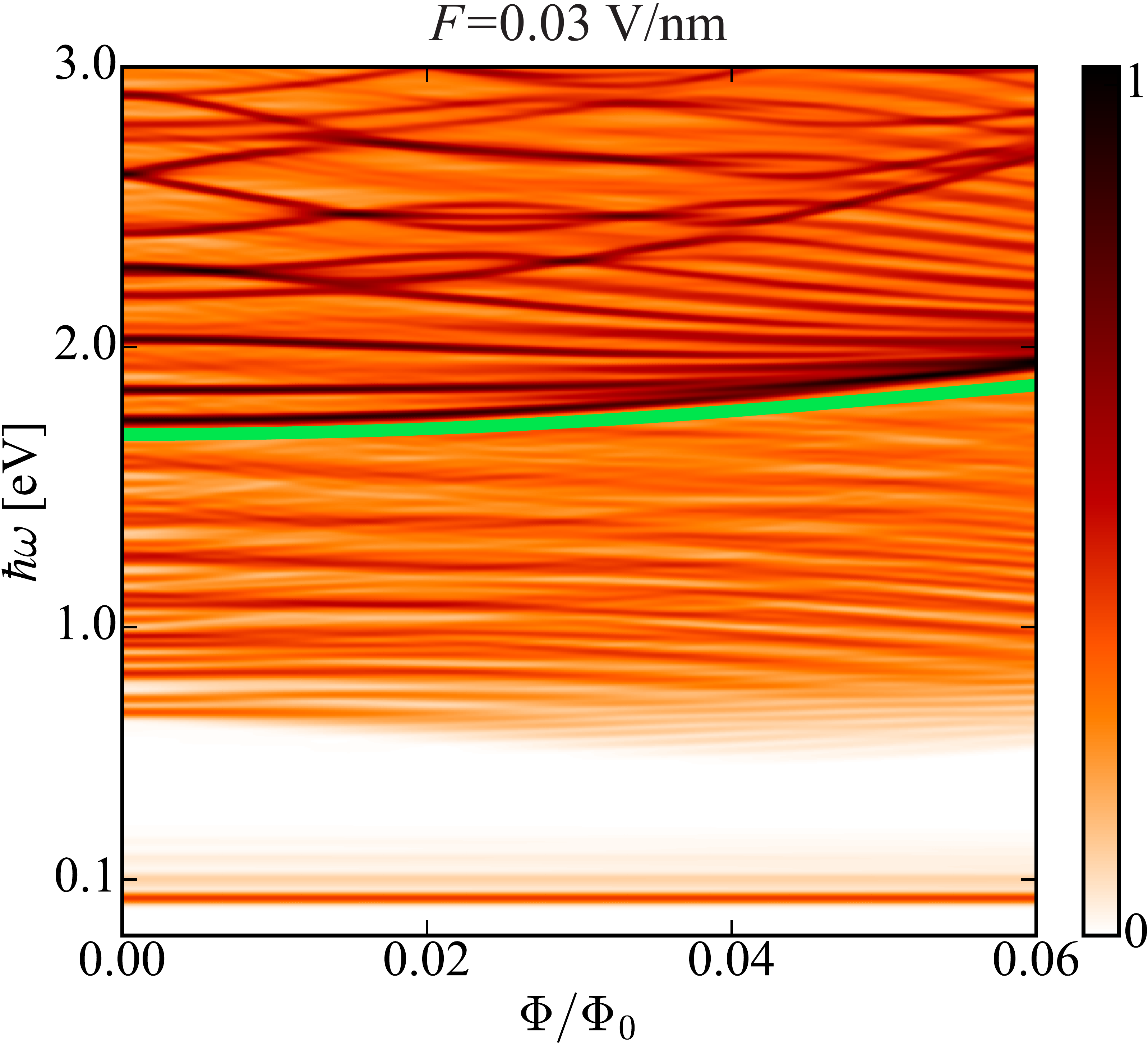}
\caption{Contour plot of the normalized optical absorption spectrum of the same RPQD as in Fig. 2 in the presence of both perpendicular magnetic field and in-plane electric field (applied along the $x$ direction). Here, we sweep the magnetic field strength and fix the electric field strength $F=0.03$ V/nm which is chosen above the critical value $F_c\simeq0.001$ V/nm in order to make the edge-to-edge transitions observable. The green solid line represents the optical gap of the RPQD. The white and black colors stand for the lowest and highest absorption intensities, respectively.} \label{Fig1}
\end{center}
\end{figure}

We also examine how the electric and magnetic fields interplay and influence the optical absorption of the RPQD. In Fig. 6, we show the contour plot of $A_{ave}(\hbar\omega)$ of the same RPQD as in Fig. 2 in the presence of both electric and magnetic fields. Here, we sweep the magnetic field strength and fix the electric field strength $F=0.03$ V/nm which is chosen above the critical value $F_c\simeq0.001$ V/nm in order to make the edge-to-edge transitions observable. Note that the optical absorption intensity is normalized with respect to its maximum value. As can be seen, unlike that induced by the edge-to-bulk and bulk-to-bulk transitions, the optical absorption induced by the edge-to-edge transitions is almost unaffected by the magnetic field due to the strong localized nature of the edge states. It should be noted that free-carrier absorption in semiconductor systems also occurs at lower photon energies, which is similar to the edge-to-edge absorption considered here. However, free-carrier absorption normally occurs in doped systems, and more importantly it changes significantly with the magnetic field due to the intra-Landau-level transitions. The robust optical absorption induced by the edge-to-edge transitions can be utilized to identify the edge states in the RPQD.

\begin{figure}[htbp]
\centering
\includegraphics[width=8.9cm]{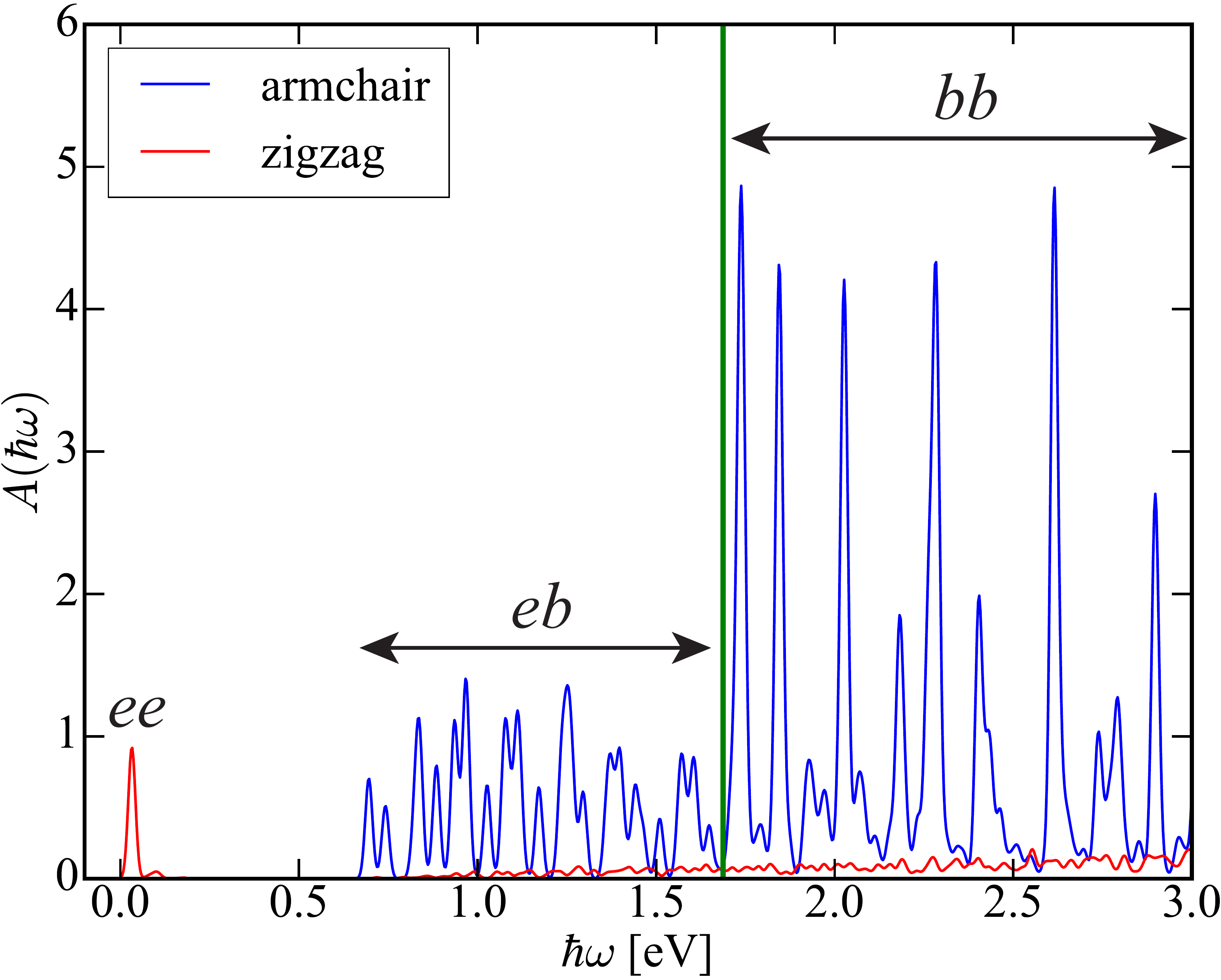}
\caption{Optical absorption spectra (arbitrary units) of the same RPQD as in Fig. 2 at a finite electric electric $F=0.03$ V/nm for different light polarizations as indicated. Here, the green solid line represents the band gap energy of the RPQD, and the characters $bb$, $eb$, and $ee$ denote the bulk-to-bulk, edge-to-bulk, and edge-to-edge transitions, respectively.} \label{Fig1}
\end{figure}

As mentioned previously, the polarization sensitive absorption can be induced by the edge-to-bulk and bulk-to-bulk transitions in the RPQD at zero electric field (see Fig. 4). In that case, the optical absorption induced by the edge-to-bulk and bulk-to-bulk transitions is much larger for light polarization along the armchair direction than for light polarization along the zigzag edge. However, we find that at nonzero electric field there is additional polarization sensitive absorption induced by the edge-to-edge transitions. But in this case, the optical absorption induced by the edge-to-edge transitions is much smaller for light polarization along the armchair direction than for light polarization along the zigzag edge. We show these results in Fig. 7 at finite electric field ($F=0.03$ V/nm). As can be seen clearly, in addition to the edge-to-bulk and bulk-to-bulk absorptions (labeled by $eb$ and $bb$ in the figure), the edge-to-edge absorption (labeled by $ee$ in the figure) is also polarization sensitive but it is almost zero for light polarization along the armchair direction. Again, this absorption feature can be understood by looking to the squared dipole matrix element $|\bra{f}x\ket{i}|^2$ ($|\bra{f}y\ket{i}|^2$) for light polarization along the armchair (zigzag) direction, where $\ket{i}$ and $\ket{f}$ are the initial and final edge states, respectively. Our numerical results indicate that at nonzero electric field $|\bra{f}y\ket{i}|^2$ is much larger than $|\bra{f}x\ket{i}|^2$ (almost vanishing) for the edge-to-edge transitions.

\section{Concluding remarks}

We have investigated the electronic and optical properties of RPQDs with armchair and zigzag edges under in-plane electric field and perpendicular magnetic field. The energy spectra and wave functions of RPQDs are obtained by solving the tight-binding model numerically. The corresponding optical absorption spectra of RPQDs are then calculated using the energy spectra and wave functions. In our calculation, we ignore the excitonic Coulomb interaction. Our results show that edge states are formed within the band gap of the RPQD which are mainly localized at the zigzag boundaries and are well separated from the bulk states above and below the band gap. The edge and bulk states have different responses to the electric and magnetic fields, leading to distinct electric- and magnetic-field dependencies of their energy spectra, i.e., the bulk states are more affected by magnetic field while the edge states are more influenced by electric field.

When applying normal incident light, we find that the PRQD can absorb (transmit) light polarized along the armchair (zigzag) direction and a large linear dichroism is observed. In an undoped RPQD and for in-plane polarized light, due to the presence of well-defined edge states, three types of optical transitions are observed under electric and magnetic fields: the first comes from the bulk-to-bulk transitions; the second from the edge-to-bulk transitions; and the third from the edge-to-edge transitions. The electric and magnetic fields influence these three types of optical transitions differently due to the different responses of bulk and edge states to these fields. Particularly, we find that the electric field can activate the edge-to-edge transitions while the magnetic field can not. The absorption intensity induced by such transitions increases with increasing electric field, while it is almost unaffected by the magnetic field. An important result obtained in the present work is that the optical absorption spectra of RPQDs under electric and magnetic fields can be utilized to determine their band gaps and to identify their edge states.

\section*{Acknowledgments}

This work was financially supported by the China Scholarship Council
(CSC), the Flemish Science Foundation (FWO-Vl), the National Natural Science Foundation of China (Grant Nos. 11304316 and 11574319), and by the Chinese Academy of Sciences (CAS).


\section*{References}

\begin{thebibliography}{99}
\bibitem{LiL2014} Likai Li, Yijun Yu, Guo Jun Ye, Qingqin Ge, Xuedong Ou, Hua Wu, Donglai Feng, Xian Hui Chen, and Yuanbo Zhang, \href{}{Nat. Nanotech. \textbf{9}, 372 (2014).}

\bibitem{LiuH2014} Han Liu, Adam T. Neal, Zhen Zhu, Zhe Luo, Xianfan Xu, David Tomanek, and Peide D. Ye, \href{}{ACS Nano \textbf{8}, 4033 (2014).}

\bibitem{QiaoJ2014} Jingsi Qiao, Xianghua Kong, Zhi-Xin Hu, Feng Yang, and Wei Ji, \href{}{Nat. Commun. \textbf{5}, 4475 (2014).}
\bibitem{XiaF2014} Fengnian Xia, Han Wang, and Yichen Jia, \href{}{Nat. Commun. \textbf{5}, 4458 (2014).}

\bibitem{NovoselovK2005} K. S. Novoselov, A. K. Geim, S. V. Morozov, D. Jiang, M. I. Katsnelson, I. V. Grigorieva, S. V. Dubonos, and A. A. Firsov, \href{}{Nature \textbf{438}, 197 (2005).}

\bibitem{WangQ2012} Qing Hua Wang, Kourosh Kalantar-Zadeh, Andras Kis, Jonathan N. Coleman, and Michael S. Strano, \href{}{Nat. Nanotech. \textbf{7}, 699 (2012).}

\bibitem{DasS2014} Saptarshi Das, Wei Zhang, Marcel Demarteau, Axel Hoffmann, Madan Dubey, and Andreas Roelofs, \href{}{Nano Lett. \textbf{14}, 5733 (2014).}

\bibitem{RodinA2014} A. S. Rodin, A. Carvalho, and A. H. Castro Neto, \href{}{Phys. Rev. Lett. \textbf{112}, 176801 (2014).}
\bibitem{WeiQ2014} Qun Wei and Xihong Peng, \href{}{Appl. Phys. Lett. \textbf{104}, 251915 (2014).}

\bibitem{LowT2014} Tony Low, A. S. Rodin, A. Carvalho, Yongjin Jiang, Han Wang, Fengnian Xia, and A. H. Castro Neto, \href{}{Phys. Rev. B \textbf{90}, 075434 (2014).}

\bibitem{FeiR2014} Ruixiang Fei, Alireza Faghaninia, Ryan Soklaski, Jia-An Yan, Cynthia Lo, and Li Yang, \href{}{Nano Lett. \textbf{14}, 6393 (2014).}

\bibitem{WangX2015} Xiaomu Wang, Aaron M. Jones, Kyle L. Seyler, Vy Tran, Yichen Jia, Huan Zhao, HanWang, Li Yang, Xiaodong Xu, and Fengnian Xia, \href{}{Nat. Nanotech. \textbf{10}, 517 (2015).}

\bibitem{LuS2015} S. B. Lu, L. L. Miao, Z. N. Guo, X. Qi, C. J. Zhao, H. Zhang, S. C. Wen, D. Y. Tang, and D. Y. Fan, \href{}{Optics Express \textbf{23}, 11183 (2015).}

\bibitem{TahirM2015} M. Tahir, P. Vasilopoulos, and F. M. Peeters, \href{}{Phys. Rev. B \textbf{92}, 045420 (2015).}

\bibitem{LiL2016} Likai Li, Fangyuan Yang, Guo Jun Ye, Zuocheng Zhang, Zengwei Zhu, Wenkai Lou, Xiaoying Zhou, Liang Li, Kenji Watanabe, Takashi Taniguchi, Kai Chang, Yayu Wang, Xian Hui Chen, and Yuanbo Zhang, \href{}{Nat. Nanotech. \textbf{11}, 593 (2016).}

\bibitem{CarvalhoA2014} A. Carvalho, A. S. Rodin, and A. H. Castro Neto, \href{}{Europhys. Lett. \textbf{108}, 47005 (2014).}

\bibitem{PengX2014} Xihong Peng, Andrew Copple, and Qun Wei, \href{}{J. Appl. Phys. \textbf{116}, 144301 (2014).}

\bibitem{TranV2014} Vy Tran and Li Yang, \href{}{Phys. Rev. B \textbf{89}, 245407 (2014).}

\bibitem{ZhangJ2014} J. Zhang, H. J. Liu, L. Cheng, J. Wei, J. H. Liang, D. D. Fan, J. Shi, X. F. Tang, and Q. J. Zhang, \href{}{Sci. Rep. \textbf{4}, 6452 (2014).}

\bibitem{EzawaM2014} Motohiko Ezawa, \href{}{New J. Phys. \textbf{16}, 115004 (2014).}

\bibitem{WuQ2015} Qingyun Wu, Lei Shen, Ming Yang, Yongqing Cai, Zhigao Huang, and Yuan Ping Feng, \href{}{Phys. Rev. B \textbf{92}, 035436 (2015).}

\bibitem{SisakhtE2015} Esmaeil Taghizadeh Sisakht, Mohammad H. Zare, and Farhad Fazileh, \href{}{Phys. Rev. B \textbf{91}, 085409 (2015).}

\bibitem{OstahieB2016} B. Ostahie and A. Aldea, \href{}{Phys. Rev. B \textbf{93}, 075408 (2016).}

\bibitem{ZhangX2015} Xiao Zhang, Haiming Xie, Zhengdong Liu, Chaoliang Tan, Zhimin Luo, Hai Li, Jiadan Lin, Liqun Sun, Wei Chen, Zhichuan Xu, Linghai Xie, Wei Huang, and Hua Zhang, \href{}{Angew. Chem. Int. Ed. \textbf{54}, 3653 (2015).}

\bibitem{SunZ2015} Zhengbo Sun, Hanhan Xie, Siying Tang, Xue-Feng Yu, Zhinan Guo, Jundong Shao, Han Zhang, Hao Huang, Huaiyu Wang, and Paul K. Chu, \href{}{Angew. Chem. \textbf{127}, 11688 (2015).}

\bibitem{ZhangR2015} Rui Zhang, X. Y. Zhou, D. Zhang, W. K. Lou, F Zhai, and Kai Chang, \href{}{2D Materials \textbf{2}, 045012 (2015).}

\bibitem{NiuX2016} XiangHong Niu, Yunhai Li, Huabing Shu, and Jinlan Wang, \href{}{J. Phys. Chem. Lett. \textbf{7}, 370 (2016).}

\bibitem{DasP2016} Paul Masih Das, Gopinath Danda, Andrew Cupo, William M. Parkin, Liangbo Liang, Neerav Kharche, Xi Ling, Shengxi Huang, Mildred S. Dresselhaus, Vincent Meunier, and Marija Drndic, \href{}{ACS Nano \textbf{10}, 5687 (2016).}







\bibitem{ZareniaM2011} M. Zarenia, A. Chaves, G. A. Farias, and F. M. Peeters, \href{}{Phys. Rev. B \textbf{84}, 245403 (2011).}

\bibitem{RudenkoA2014} A. N. Rudenko and M. I. Katsnelson, \href{}{Phys. Rev. B \textbf{89}, 201408(R) (2014).}

\bibitem{MoldovanD2016} D. Moldovan and F. M. Peeters, \textit{Pybinding v0.8.0: a Python package for tight-binding calculations}, \href{}{http: //dx.doi.org/10.5281/zenodo.56818.}



\bibitem{LeeS2002} S. Lee, J. Kim, L. Jonsson, J. W. Wilkins, G. W. Bryant, and G. Klimeck, \href{}{Phys. Rev. B \textbf{66}, 235307 (2002).}

\bibitem{AbdelsalamH2016} Hazem Abdelsalam, Mohamed H. Talaat, Igor Lukyanchuk, M. E. Portnoi, and V. A. Saroka, \href{}{J. Appl.
    Phys. \textbf{120}, 014304 (2016).}

\bibitem{Son2006} Young-Woo Son, Marvin L. Cohen, and Steven G. Louie, \href{}{Phys. Rev. Lett. \textbf{97}, 216803 (2006).}

\bibitem{Bollinger2001} M. V. Bollinger, J. V. Lauritsen, K. W. Jacobsen, J. K. N{\o}rskov, S. Helveg, and F. Besenbacher, \href{}{Phys. Rev. Lett. \textbf{87}, 196803 (2001).}

\bibitem{HuangY2008} Y. C. Huang, M. F. Lin, and C. P. Chang, \href{}{J. Appl. Phys. \textbf{103}, 073709 (2008).}

\bibitem{KoshinoM2008} Mikito Koshino and Tsuneya Ando, \href{}{Phys. Rev. B \textbf{77}, 115313 (2008).}

\bibitem{TabertC2013} C. J. Tabert and E. J. Nicol, \href{}{Phys. Rev. B \textbf{88}, 085434 (2013).}

\bibitem{YuanS2015} S. Yuan, A. N. Rudenko, and M. I. Katsnelson, \href{}{Phys. Rev. B \textbf{91}, 115436 (2015).}

\end{thebibliography}
\end{document}